\documentclass[runningheads]{llncs}

\usepackage{graphicx}
\usepackage{booktabs}
\usepackage{multicol}
\usepackage{multirow}
\usepackage{pgfplots}
\usepackage[utf8]{inputenc}
\usepackage{csvsimple}
\usepackage{tikz}
\usepackage{pbox}
\usepackage[margin=1.5in]{geometry}
\usetikzlibrary{pgfplots.groupplots}

\usepackage{amssymb,amsmath,amsthm,enumitem}
\usepackage{subcaption}
\usepackage{lipsum}

\begin{document}
\title{An Axiomatic Study of Query Terms Order in Ad-hoc Retrieval}
%
%
 \author{Ayyoob Imani\inst{1} \and
 Amir Vakili\inst{1} \and
Ali Montazer\inst{2} \and Azadeh Shakery\inst{1}}
%
%
 \institute{Tehran University, Tehran, Iran \email{ \{ayyoub.imani, shakery, a\_vakili\} @ut.ac.ir}
 \and University of Massachusetts Amherst, Amherst, USA \email{montazer@umass.edu}}
%
\maketitle              
\begin{abstract}
Classic retrieval methods use simple bag-of-word representations for queries and documents. This representation fails to capture the full semantic richness of queries and documents. More recent retrieval models have tried to overcome this deficiency by using approaches such as incorporating dependencies between query terms, using bi-gram representations of documents, proximity heuristics, and passage retrieval.
While some of these previous works have implicitly accounted for term order, to the best of our knowledge, term order has not been the primary focus of any research. In this paper, we focus solely on the effect of term order in information retrieval. We will show that
documents that have two query terms in the same order as in the query have a higher probability of being relevant than documents that have two query terms in the reverse order. Using the axiomatic framework for information retrieval, we introduce a constraint that retrieval models must adhere to in order to effectively utilize term order dependency among query terms.
We modify existing retrieval models based on this constraint so that if the order of a pair of query terms is semantically important, a document that includes these query terms in the same order as the query should receive a higher score compared to a document that includes them in the reverse order.
Our empirical evaluation using both TREC newswire and web corpora demonstrates that the modified retrieval models significantly outperform their original counterparts.

\end{abstract}

\section{Introduction}
Classic information retrieval systems such as BM25 \cite{Robertson} or query likelihood \cite{Ponte} use a very simple bag of word representations for both queries and documents. These models have proven to be effective and offer a compromise between efficiency and good results. However, query terms have associations that are not considered when using a bag of word representation and this causes a decrease in performance of classic information retrieval systems. Recent research has shown that taking these associations into consideration can effectively improve retrieval performance \cite{bendersky2010learning,peng2007incorporating,lv2009positional,huston2014comparison,bendersky2012modeling,yu1983generalized}.

Positional language models capture proximity and passage retrieval heuristics for information retrieval \cite{lv2009positional}. The method proposed by Yu attempts to incorporate dependencies between terms using term co-occurrences information of terms \cite{yu1983generalized}. While these two previous works on terms associations totally neglect the role of term order, methods that use n-grams \cite{bendersky2012modeling,peng2007incorporating,bendersky2010learning,metzler2005markov}, implicitly consider the order of terms for adjacent terms to some extent but don't consider the order of terms that are separated by a few other terms. These methods suffer from data sparsity and using a bigger n-gram to cover this small distance causes even more sparsity. An important difference between these methods and our proposed method is that we consider order dependency not only between two adjacent terms but also for more distant terms inside a specified window size thus solving the data sparsity problem.


In this paper, we hypothesize that if a query contains term pairs whose semantics changes if they appear in reverse, documents where these terms appear in the same order, are more likely to be relevant. To verify the hypothesis, we conduct data exploratory analysis using various TREC collections.

We use the axiomatic framework proposed by Fang \cite{fang2004formal} to model this hypothesis as a formal constraint. We determine that SDM \cite{metzler2005markov} and PLM do not satisfy this constraint and then we modify these two retrieval methods so they adhere to this constraint. Experiments show that our modified models significantly outperform the baselines.


\section{Methodology}
Axiomatic analysis provides an approach for developing retrieval models based on formalized constraints and has received much attention in the information retrieval community \cite{Montazer,fang2004formal}. 
In this section, we explain the intuition behind our term order constraint before formally defining it. Finally, we propose two methods for modifying SDM and PLM retrieval methods.

\begin{table}
\centering
\caption{Association of relevance and query terms order along with collection statistics}
\small
\begin{tabular}{llccccc}
     \toprule
     & Collection & Queries                   & \#docs & \(p(rel|q_1q_2)\) & \(p(rel|q_2q_1)\) \\ 
     \midrule
     AP & Associated Press 1988-89 &  51-200             & 165k& 0.212 & 0.192\\
     Robust & TREC 2004 Robust Track&  301-450, 601-700 & 528k & 0.194 & 0.167  \\
     WT2g & TREC 1999 Web Track &  401-450            &  247k & 0.191 & 0.171  \\
     WT10g &TREC 2000-2001 Web Track &  451-550 			& 1.7M 		& 0.116 & 0.096 \\
     \bottomrule
\end{tabular}
\label{table:prel}
\end{table}

\subsection{Term order and document relevance}
In this section we check whether our intuition regarding the effect of term order is correct. To achieve this, we compute \(p(Rel| \text{ordered match})\) and \(p(Rel | \text{reversed match})\) and test whether \(p(Rel| \text{ordered match})\) is significantly higher than \(p(Rel | \text{reversed match})\). 

For all queries in the dataset we find term pairs \(q_1 q_2\) where \(q_1\) comes before \(q_2\) in a window of size 5. Having relevance judgments for the queries we compute the following 
\begin{align*}
p(Rel| \text{ordered match}) &= p(Rel|q_1q_2) = \frac{Rdf(q_1q_2)}{Rdf(q_1q_2) + Rdf(q_2q_1)} \\
p(Rel| \text{reversed match}) &= p(Rel|q_2q_1) = \frac{Rdf(q_2q_1)}{Rdf(q_1q_2) + Rdf(q_2q_1)}
\end{align*}
where \(Rdf(q_1q_2)\) is the relevant document frequency of the two terms \(q_1\) and \(q_2\) appearing	 in order inside a window of size 5.
Table \ref{table:prel} summarizes the results. The results suggest the probability of relevance for a document having the terms in the same order as query is more than a document that has them in reverse order. For all four datasets, the difference is statistically significant using the two tailed paired t-test computed at a 95\% confidence level.

\subsection{Definition of the query term order constraint}
This constraint is defined to capture term ordering in documents and queries which is lost in existing retrieval models. If the semantics of a pair of terms in a query differs when their ordering is reversed, this constraint will ensure that a document with these terms in the correct order will have a higher relevance score than a document which has them in the reverse order.

Formally, let \(D = \langle w_1, \cdots, w_m \rangle \) be a document where \(w_i\) is the term at position \(i\)
and \(Q = \langle q_1, q_2 \rangle \) be a query with two terms \(q_1\) and \(q_2\) such that \(
    sem( q_1 \ q_2 ) \neq sem( q_2 \ q_1 )
\)
where \(sem( q_1 \ q_2 )\) denotes the semantic meaning of the phrase ``\(q_1 \ q_2\)". When the above equation is true it indicates that the semantic meaning of the phrase ``\(q_1\) \(q_2\)" is not the same as ``\(q_2\) \(q_1\)". 

We then define \(D_1 = D \Vert \langle q_1 q_2 \rangle = \langle d_1, \cdots, d_m, q_1, q_2 \rangle \) and \( D_2 = D \Vert \langle q_2, q_1 \rangle  = \langle d_1, \cdots, d_m, q_2, q_1 \rangle \) where \(\Vert\) is the notation for concatenation (e.g. \(D_1\) is a document created from inserting the terms \(q_1\) and \(q_2\) in that order to the end of \(D\)). Then, we can say
\(
    S(D_2, Q) \leq S(D_1, Q)
\)
where \(S(D,Q)\) denotes the relevance score of document \(D\) with respect to query \(Q\). Based on this constraint, we want the retrieval function to give a higher score to a document which has the two query terms in the same order as the query .

\subsection{Modification of existing retrieval methods}
To the best of our knowledge, existing retrieval models such as bag-of-words, n-gram based, and passage retrieval models do not satisfy the proposed constraint. In this section, we select SDM \cite{metzler2005markov} and PLM \cite{lv2009positional} as examples of n-gram retrieval and robust passage retrieval models respectively, and then modify them so as to satisfy the proposed term order constraint. Modification of other retrieval models such as BM25 and query likelihood are also possible as they are simpler compared to the chosen models.

To introduce our modification to existing methods, we introduce the following notation:
\textbf{Df(w,w')} is the frequency of documents that contain terms \(w\) and \(w'\) in this order in a window of specific size.

The proposed constraint imposes a stipulation that we should only consider term order for terms whose order is semantically important so we need to define a function that captures whether the order of two query terms is important or not. For this purpose we define:

\begin{equation} \label{eq:SITO}
sem(w,w')=\left|1/2 - \frac{Df(w,w')}{Df(w,w') + Df(w',w)}\right|
\end{equation}

The above function captures the importance of term order for two query terms. This function ranges from 0 to 1/2. When the difference between document frequency of \(w w'\) and \(w' w\) is large, we can conclude that different orders of these two terms are pointing to different concepts and the \(sem\) function will evaluate to a value close to 1/2. But when the difference between document frequencies for the different orders of these two terms is not high we do not have enough evidence to decide with certainty whether the different orders are pointing to different concepts and the \(sem\) function evaluates to 0. 

While the proposed function is rather simple and computationally efficient, it gives satisfactory results. We call this function semantic importance of term order (SITO).

\subsubsection{Modification of SDM}
The Sequential dependency model is a retrieval function that incorporates both term bigrams and term proximity. The score of a document \(D\) with respect to query \(Q\) is calculated as:
\begin{gather}
P(D|Q) = \lambda_T\sum_{q\in Q}f_T(q,D) + \lambda_O\sum_{q_i,q_{i+1}\in Q}f_O(q_i,q_{i+1},D) + \lambda_U \sum_{q_i,q_{i+1}\in Q} f_U(q_i,q_{i+1},D) 
\end{gather}

where $\lambda_T$, $\lambda_O$ and $\lambda_U$ are hyper-parameters dictating the importance of unigram frequency, ordered bigram frequency and unordered term co-occurrence frequency within a window which are defined below:
\begin{align*}
& f_O(q_i,q_{i+1},D)=\log\left[\frac{tf_{\#1(q_i,q_{i+1},D}+\mu \frac{cf_{\#1(q_i,q_{i+1}}}{|C|}}{|D|+\mu}\right] && \pbox{10cm}{Weight of exact phrase \\ ``$q_i\ q_{i+1}$" in doc $D$} \\
&f_U(q_i,q_{i+1},D)=\log\left[\frac{tf_{\#uw8(q_i,q_{i+1},D}+\mu \frac{cf_{\#uw8(q_i,q_{i+1}}}{|C|}}{|D|+\mu}\right] && \pbox{10cm}{weight of unordered (span=8)\\ window ``$q_i\ q_{i+1}$" in $D$}  
\end{align*}




In order for SDM to satisfy the proposed constraint, it should take into account not just sequential term pairs, but all terms appearing together within a window. Therefore we add a component to SDM which calculates ordered term co-occurance. The function should also take into account the semantic importance of word order (SITO) when rewarding terms appearing in order. 
The modified SDM function is as follows:
\begin{align*}
P(D|Q) =& \lambda_T\sum_{q\in Q}f_T(q,D) + \lambda_O\sum_{q_i,q_{i+1}\in Q}f_O(q_i,q_{i+1},D)g(q_{i},q_{i+1}) + \\ & \lambda_U \sum_{q_i,q_{i+1}\in Q} f_U(q_i,q_{i+1},D)h(q_{i},q_{i+1})
+ \sum_{q_i,q_{j}\in Q,i+1<j}\lambda_{OW}f_{OW}(q_i, q_j, D)g(q_{i},q_{j})
\end{align*}
where \(g(w_1,w_2)=\frac{3}{4} + sem(w_1,w_2)\) and \(h(w_1,w_2)=\frac{5}{4} - sem(w_1,w_2)\). 
We define them as such since \(sem\) has a range of \([0,5]\) and larger values indicate term order is semantically important.
\(g(\cdot,\cdot)\) increases or decreases the weight based on whether term order is semantically important or not. \(h(\cdot,\cdot)\) does the opposite.  \(\lambda_{OW}\) is the weight we would like to give to the ordered co-occurrence component. \(f_{OW}\) is defined as  

\begin{align*}
f_{OW}(q_i,q_{j},D)=\log\left[\frac{tf_{\#owN(q_i,q_{j},D)}+\mu \frac{cf_{\#owN(q_i,q_{j})}}{|C|}}{|D|+\mu}\right] && \pbox{10cm}{weight of ordered (span=N)\\ window ``$q_i$", ``$q_{j}$" in $D$}  
\end{align*}

A similar approach can be taken for modifying the full dependency model (FDM) \cite{metzler2005markov} and the weighted sequential dependency model (WSDM) \cite{bendersky2010learning}.

\subsubsection{Modification of the PLM model}

Before we introduce our modification to PLM, we provide a short overview of the model. Let \(D = (w_1, w_2, \cdots, w_N)\) be a document of size \(N\) where \(w_i\) shows the \(i\)-th term of the document. Let \(c(w, j)\) be the count of term \(w\) at position i in document \(D\) (if \(w\) occurs at position \(i\), it is 1, otherwise 0) and \(k(i, j)\) be the propagated count to position \(i\) from a term at position \(j\). PLM defines the total propagated count of term \(w\) at position \(i\) from the occurrences of \(w\) in all the positions as:

\[c'(w, i) = \sum_{j=1}^N c(w, j)k(i,j)\]

Based on this term propagation, PLM has a frequency vector \\
\(\langle c'(w_1,i), c'(w_2,i), \cdots, c' (w_N,i)\rangle\) at position i forming a virtual document \(D'_i\). PLM then uses the language modeling approach for information retrieval and computes the score of document \(D'_i\) using KL\_divergence retrieval model. Finally, PLM calculates the overall score of \(D\) based on the scores of these virtual documents.

In order for PLM to satisfy our word order constraint, we need to reward documents in which matched query terms appear in order with some other query terms in the document, therefore if a term in position \(i\) appears in order with another query term, we increase the score the document receives from this term. To achieve this we multiply \(c(w,i)\) with a weight that captures the semantic importance of term order. This will ensure that the score a document receives from a term will increase if the term appears in order with some other query terms.

\[
   c'(w, i, D, Q) = \sum_{j=1}^N c(w, j)k(i,j)\text{weight}(w_j,D, Q) 
\]

If the term at position \(j\) is not in order with any other query terms around position \(j\), this weight will be 1, but if another term appears in order with this query term around position \(j\), the weight will be increased proportionally to the  semantic importance of these two terms (equation \ref{eq:SITO}). We define the weight function as follows:

\[
    \text{weight}(w_j, D, Q) = 1 + \sum_{w' \in Q} \lambda  \cdot sem(w_j, w') \cdot I(w_j, w', D, Q)
\]
where \(I(w_j, w', D, Q)\) is true if \(w_j\) and \(w'\) appear together in the same order they appear in the query within a specified window size around position \(j\) and \(\lambda\) is a free parameter to control to what extent the weight function affects a term's score.


  

The modifications proposed in this section has the effect of rewarding query terms appearing in order in documents and therefore satisfying the term order constraint.



\begin{figure}
	\pgfplotscreateplotcyclelist{grayscale}{
    thick,white!20!black,mark=o,mark options=solid,densely dotted\\%
    thick,white!50!black,mark=x,mark options=solid,dashed\\%
    thick,white!50!black,mark=diamond,mark options=solid,solid\\%
    thick,white!50!black,mark=triangle,mark options=solid,solid\\%
}
    \vspace*{-5mm}
    \centering
    \small
     \begin{subfigure}{.3\textwidth}
\centering

\begin{tikzpicture}
\begin{axis}[
    y tick label style={
        /pgf/number format/.cd,
            fixed,
            fixed zerofill,
            precision=2,
        /tikz/.cd
    },
    legend style={font=\tiny},
    xlabel={Window size},
    ylabel={MAP},
    width=4cm,
    legend pos=outer north east,
    cycle list name=grayscale,
]
\addplot table [x=window size, y=A, col sep=comma] {sdm_maps7.csv};
\addplot table [x=window size, y=R, col sep=comma] {sdm_maps7.csv};
\addplot table [x=window size, y=W, col sep=comma] {sdm_maps7.csv};
\addplot table [x=window size, y=W10, col sep=comma] {sdm_maps7.csv};

\end{axis}
\end{tikzpicture}
\caption{}
\end{subfigure}
\begin{subfigure}{.33\textwidth}
\centering

\begin{tikzpicture}
\begin{axis}[
    y tick label style={
        /pgf/number format/.cd,
            fixed,
            fixed zerofill,
            precision=2,
        /tikz/.cd
    },
    legend style={font=\tiny},
    ytick={0.24,0.28,0.32},
    xlabel={Window size},
    width=4cm,
    legend pos=outer north east,
    cycle list name=grayscale,
]
\addplot table [x=window size, y=A, col sep=comma] {plm_maps6.csv};
\addplot table [x=window size, y=R, col sep=comma] {plm_maps6.csv};
\addplot table [x=window size, y=W, col sep=comma] {plm_maps6.csv};
\addplot table [x=window size, y=W10, col sep=comma] {plm_maps6.csv};

\end{axis}
\end{tikzpicture}
\caption{}
\end{subfigure}
\begin{subfigure}{.33\textwidth}
\centering

\begin{tikzpicture}
\begin{axis}[
    y tick label style={
        /pgf/number format/.cd,
            fixed,
            fixed zerofill,
            precision=2,
        /tikz/.cd
    },
    xtick={0.5, 1, 2, 4, 8},
    ytick={0.22, 0.26, 0.30, 0.34},
    legend style={font=\tiny},
    xlabel={\(\lambda\)},
    width=4cm,
    legend pos=outer north east,
    xmode=log,
    log ticks with fixed point,
    cycle list name=grayscale,
]
\addplot table [x=lambda, y=A, col sep=comma] {lambda6.csv};
\addplot table [x=lambda, y=R, col sep=comma] {lambda6.csv};
\addplot table [x=lambda, y=W, col sep=comma] {lambda6.csv};
\addplot table [x=lambda, y=W10, col sep=comma] {lambda6.csv};
\legend{AP, Robust, WT2G, WT10G}
\end{axis}
\end{tikzpicture}
\caption{}
\end{subfigure}
    \caption{Figure (a) shows the effect of window size on MAP for SDM-M, figure (b) shows the effect of window size on MAP for PLM-M at \(\lambda = 4\), figure (c) shows the effect of \(\lambda\) on MAP for PLM at a windows size of 4}
\label{fi:charts}
\end{figure}
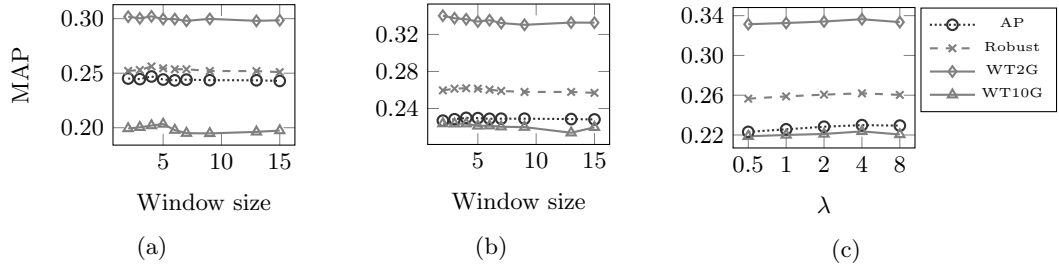

\section{Experiments}
In this section, we evaluate our proposed modifications to SDM and PLM. Our aim is to analyze and compare retrieval effectiveness of the proposed method across different collections with different features. We used four standard TREC collections in our experiment AP88-89, Robust, WT2G, and WT10G. The first two collections are news collections, and the last two are web collections with more noisy documents. The statistics for these collections are shown in table \ref{table:prel}. We take the titles of topics as queries. We stem the documents and queries using the Porter stemmer. The experiments on PLM and SDM were carried out on the Lemur toolkit and the Galago toolkit respectively as these we the tools the original authors used for their implementations\footnote{http://lemurproject.org/}.




We use mean average precision (MAP) of the top 1000 ranked documents as our evaluation metric. Statistical significance testing is performed using two-tailed paired t-test at a 95\% confidence level.


\begin{table}
\caption{Comparison of the modified retrieval methods with their baselines.}
    \centering
    \small
    \setlength\tabcolsep{1.5pt}
    \begin{tabular}{l@{\hskip .1in}ll@{\hskip .1in}ll@{\hskip .1in}ll@{\hskip .1in}ll}
        \toprule
         \multirow{2}{*}{} & \multicolumn{2}{c}{AP} &
         \multicolumn{2}{c}{Robust} & \multicolumn{2}{c}{WT2g} & \multicolumn{2}{c}{WT10g}\\ 
         \cmidrule(lr){2-3}\cmidrule(lr){4-5}\cmidrule(lr){6-7}\cmidrule(lr){8-9}
                 &  MAP  & P@10  & MAP   & P@10   & MAP   & P@10  & MAP   & P@10  \\
         \midrule
         SDM    & 0.2358  & 0.3483  & 0.2472  & 0.4149  & 0.2902  & 0.4300  & 0.1951 & 0.2369\\
         SDM-M  & 0.2446* & 0.3839* & 0.2562* & 0.4197* & 0.3021* & 0.4260 & 0.2034* & 0.2500*\\
         \midrule
         PLM     & 0.2198  & 0.3483  & 0.2538  & 0.4305  & 0.3287  & 0.4520  & 0.2073 & 0.2640\\
         PLM-M   & 0.2299* & 0.3678* & 0.2619* & 0.4378 & 0.3364* & 0.4520 & 0.2236* & 0.2560\\
         \bottomrule
    \end{tabular}
    \label{table:results}
\end{table}

\subsection{Evaluation of modified methods}


We compare each modified method with its unmodified counterpart as the baseline. The results are summarized in table \ref{table:results}. Modified methods result in a statistically significant improvement for all four datasets. The modifications have a greater effect on the WT2G and WT10g datasets. This is most likely due to the fact that AP88-89 and Robust are homogeneous collections but WT2G and WT10G are heterogeneous and therefore noisier. As reported previously in \cite{lv2009positional} term dependency information is more helpful on noisy datasets.

We ran the experiment with window sizes between 2 and 15. Figures \ref{fi:charts}.a and \ref{fi:charts}.b shows the sensitivity of MAP to the window size parameter for SDM-M and PLM-M. The best results are achieved at a window size of around 4. This is expected as term order between distant terms is meaningless and small windows sizes fail to detect semantic importance between all terms and therefore lose some information.

Figure \ref{fi:charts}.c  shows the sensitivity of the modified PLM method to parameter \(\lambda\). Increasing this parameter to large numbers increases document scores by an unreasonable amount and if we choose a very small value for this parameter, changes to document score will be ineffective. The best choice for all four datasets is to set this parameter to 4. To see whether different values of \( \lambda \) may affect window size, for each window size we further compared the results of different values of \( \lambda \) and observed that the effect of \(\lambda\) on MAP is unaffected by window size, and the best choice for \(\lambda\) for any window size is still 4.

\section{Conclusions}
In this paper we used the axiomatic framework to propose a query term order constraint for ad-hoc retrieval which states that if the order of two query terms is semantically important, a document that has these two terms in the same order as the query should get a higher score than a document that has them in the reverse order. Furthermore, we proposed modifications to two well-known and robust information retrieval methods (SDM and PLM) so as to satisfy the proposed constraint. The modifications make use of term order information to effectively improve the performance of the baselines. 
Experimental results show the proposed modifications cause a significant improvement over the baselines and a window of size 4 is the best choice for considering term order dependency.

One future research direction is to search for a better SITO function and a more integrated way to make state-of-the-art retrieval methods satisfy the proposed constraint.

%
%
%
\bibliographystyle{splncs04}
\bibliography{samplepaper}
%




\end{document}